\RequirePackage{lineno}
\documentclass[aps,prl,twocolumn,showpacs,groupedaddress]{revtex4}  
\usepackage{graphicx}  
\usepackage{dcolumn}   
\usepackage{bm}        
\usepackage{amssymb}   
\usepackage{subfigure}

\begin{document}

\hspace{5.2in} \mbox{Fermilab-Pub-10-075-E}

\title{Search for Randall-Sundrum gravitons in the dielectron and diphoton final states with 5.4 $\rm {\bf{fb^{-1}}}$ of data from $\bf{p\bar{p}}$ collisions at $\bf{\sqrt{s}=1.96}$ TeV}
%
%
\author{V.M.~Abazov$^{36}$}
\author{B.~Abbott$^{74}$}
\author{M.~Abolins$^{63}$}
\author{B.S.~Acharya$^{29}$}
\author{M.~Adams$^{49}$}
\author{T.~Adams$^{47}$}
\author{E.~Aguilo$^{6}$}
\author{G.D.~Alexeev$^{36}$}
\author{G.~Alkhazov$^{40}$}
\author{A.~Alton$^{62,a}$}
% ...

\affiliation{\vspace{0.1 in}(The D\O\ Collaboration)\vspace{0.1 in}}
% ...
\affiliation{$^{6}$Simon Fraser University, Burnaby, British Columbia, Canada;
                and York University, Toronto, Ontario, Canada}
% ...
\affiliation{$^{29}$Tata Institute of Fundamental Research, Mumbai, India}
% ...
\affiliation{$^{36}$Joint Institute for Nuclear Research, Dubna, Russia}
% ...
\affiliation{$^{40}$Petersburg Nuclear Physics Institute,
                St. Petersburg, Russia}
% ...
\affiliation{$^{47}$Florida State University, Tallahassee, Florida 32306, USA}
% ...
\affiliation{$^{49}$University of Illinois at Chicago,
                Chicago, Illinois 60607, USA}
% ...
\affiliation{$^{62}$University of Michigan, Ann Arbor, Michigan 48109, USA}
\affiliation{$^{63}$Michigan State University,
                East Lansing, Michigan 48824, USA}
% ...
\affiliation{$^{74}$University of Oklahoma, Norman, Oklahoma 73019, USA}
% ...
\date{April 11, 2010}

\begin{abstract}
Using 5.4 $\rm fb^{-1}$ of integrated luminosity from $p\bar{p}$ collisions at $\sqrt{s}=1.96$ TeV collected by the D0
detector at the Fermilab Tevatron Collider, we search for decays of the lightest Kaluza-Klein mode of the 
graviton in the Randall-Sundrum model to $ee$ and $\gamma\gamma$. 
We set 95\% C.L. lower limits on the mass of the lightest graviton between $\rm 560$ and $1050$~GeV 
for values of the coupling $k/\bar{M}_{\rm Pl}$ between 0.01 and 0.1. 

\end{abstract}

\maketitle 
\newpage

The large disparity between the scale of quantum gravity, i.e., the Planck scale, $M_{\rm Pl}\approx 10^{16}$ TeV, 
and the
electroweak scale, of the order of 1 TeV, is known in the standard model (SM) as
the hierarchy problem. In the presence of this hierarchy of scales it is not
possible to stabilize the Higgs boson mass at the low values required by experimental
data, unless by using an unlikely large amount of fine-tuning.

In the Randall-Sundrum model~\cite{rs}, the existence of a fifth dimension with a warped spacetime
metric is proposed, bounded by two three-dimensional branes. 
The SM fields are localized on one brane, while gravity originates on the other. With this configuration, 
TeV scales are naturally generated from the Planck scale due to a 
geometrical exponential factor (the ``warp factor''),
$\Lambda_{\pi}=\bar{M}_{\rm Pl}{\rm exp}(-k\pi r_{c})$, if $kr_{c}\approx 12$, where 
$\bar{M}_{\rm Pl}=M_{\rm Pl}/\sqrt{8\pi}$ is the reduced Planck scale, 
and $k$ and $r_{c}$ are the curvature scale and compactification radius of the 
extra dimension, respectively. 

Gravitons are the only particles that propagate in the fifth dimension, and appear
as a Kaluza-Klein series~\cite{kk} of massive excitations (KK gravitons, $G$) 
with spin 2, mass splitting of the order of 1 TeV, and a universal coupling to the SM fields. 
Phenomenologically, it is convenient to express the two Randall-Sundrum parameters $k$ and $r_{c}$ in terms 
of two direct observables: the mass of the lightest excitation, $M_{\rm 1}$, and 
the dimensionless coupling to the SM fields, $k/\bar{M}_{\rm Pl}$.
To address the hierarchy problem without the need for fine-tuning, $M_{\rm 1}$ should be in 
the TeV range and $0.01 \leq k/\bar{M}_{\rm Pl} \leq 0.1$~\cite{DHR}. 
KK graviton resonances could be
produced in high energy particle collisions and would subsequently decay to pairs
of SM fermions or bosons.

In this Letter, we report an inclusive search for the lightest KK graviton in the $ee$ and 
$\gamma\gamma$ decay channels with the D0 detector at the Fermilab Tevatron Collider, where protons 
and antiprotons collide at $\sqrt{s}=1.96$ TeV. KK gravitons would be produced via quark-antiquark
annihilation and gluon-gluon fusion processes. For $k/\bar{M}_{\rm Pl}\leq 0.1$, KK gravitons 
would appear as narrow resonances in the $ee$ and $\gamma\gamma$ invariant mass spectra,
with a natural width much smaller than the resolution of the D0 detector and with a 
branching fraction for the $\gamma\gamma$ decay mode which is twice that of the decay to $ee$. 
Previous D0 searches for KK gravitons have excluded $M_1<300$~GeV for 
$k/\bar{M}_{\rm Pl} =0.01$ and $M_1<900$~GeV for $k/\bar{M}_{\rm Pl}=0.1$ at the 95\% 
C.L.
~\cite{d0result}. 
CDF has recently excluded $M_1<889$~GeV for 
$k/\bar{M}_{\rm Pl}=0.1$ at the 95\% C.L. 
~\cite{cdfresult}. 

The D0 detector~\cite{run2det,run1det} consists of tracking detectors, calorimeters, and
a muon spectrometer. The tracking system includes a 
silicon microstrip tracker close to the beam and a central fiber tracker, 
both located within a 2~T superconducting solenoidal magnet. The liquid-argon and uranium 
calorimeters consist of a central section covering pseudorapidities $|\eta| \lesssim 1.1$
and two end cap calorimeters that extend the coverage to $|\eta|\approx4.2$, where
$\eta = \rm -ln[tan(\theta/2)]$, and $\theta$ is the 
polar angle with respect to the proton beam direction. The azimuthal angle is denoted
by $\phi$. The electromagnetic (EM) section of the calorimeters is segmented into four
longitudinal layers (EM$i$, $i$=1,4) with transverse segmentation of 
$\Delta\eta\times \Delta\phi=0.1\times 0.1$, except for the more finely segmented EM3 section
where it is $0.05\times 0.05$. A preshower system (CPS) uses plastic scintillators
with different orientations located between the solenoid and the cryostat of the
central calorimeter and
provides precise measurements of the positions of EM showers. 
The luminosity is measured using plastic scintillator 
arrays placed in front of the end cap calorimeters. 
The data sample was
collected between July 2002 and June 2009 using triggers requiring 
at least two clusters of energy deposits in
the EM calorimeter and corresponds to an integrated luminosity of $\rm 5.4\pm0.3~fb^{-1}$. 

We select events with two EM clusters, each with transverse momentum $p_T>25$~GeV and $|\eta|<1.1$, 
reconstructed in a cone of radius $R=\sqrt{(\Delta\eta)^2+(\Delta\phi)^2}=0.4$. The EM clusters
are required to have at least 97\% of their energy deposited in the EM calorimeter and to have 
the calorimeter isolation variable $I=[E_{\rm tot}(0.4)-E_{\rm EM}(0.2)]/E_{\rm EM}(0.2)<0.07$,
where $E_{\rm tot}(R)$ [$E_{\rm EM}(R)$] is the total [EM] energy in a cone of radius $R$. 

Given the
different branching fractions for the $\gamma\gamma$ and $ee$ decays of the KK graviton, plus the 
fact that the two channels have different backgrounds, the analysis treats the two channels
separately to optimize the sensitivity. 
If both EM clusters 
in an event are spatially matched to tracker activity, either a reconstructed track or a density 
of hits in the silicon microstrip tracker and central fiber tracker consistent with that of an electron, the event goes in
the $ee$ category. Otherwise, the event is put in the $\gamma\gamma$ category, which contains events with 
at least one EM cluster failing to match tracker activity. With this definition, about 97\% of 
the selected $G \rightarrow ee$ events are put in the $ee$ category and 
about 90\% of the selected $G \rightarrow \gamma\gamma$ events are put in the 
$\gamma\gamma$ category.

In the $ee$ category, the two electrons are not required to have opposite charges to avoid the loss due to 
charge misidentification, 
and two additional requirements are placed on each 
EM cluster: (i) the scalar sum of the $p_T$ of all tracks 
originating 
from the primary vertex (PV, see below) in an annulus of $0.05<R<0.4$ around the cluster, $I_{trk}$, be less than $2.5$~GeV; 
(ii) the cluster be consistent with the electron shower shape using a $\chi^2$ test 
and a neural network discriminant~\cite{hgg}. 
In the $\gamma\gamma$ category, additional requirements are placed on each EM cluster:
(i) $I_{trk}<2.0$~GeV;
(ii) the energy-weighted shower width in the $r-\phi$ plane in 
EM3 be less than 3.7 cm; (iii) the cluster be consistent with the photon shower shape 
using a neural network discriminant. 

Proper reconstruction of the event kinematics requires correct identification of the 
PV of the hard collision.
For events in the $ee$ category, the PV is chosen from the list of vertex
candidates as the one with the least probability of being a vertex from a soft $p\bar{p}$ interaction
as estimated from the $p_T$ of associated tracks.
For the $\gamma\gamma$ category, we use the EM-CPS pointing capability, which reconstructs
the axes of EM showers by fitting straight lines to shower positions measured in the four
longitudinal calorimeter layers and the CPS. 
The EM-CPS pointing spatial resolution is $3.7\pm0.2$ cm along the beam axis. 
If at least one photon candidate is matched to a CPS cluster~\cite{gmsb}, 
the vertex consistent with the EM-CPS pointing position is chosen as the PV.
For events with no photon candidate having a CPS match or events with inconsistent EM-CPS 
pointing positions of the two photon candidates, 
the PV is chosen as the one with the highest number of associated tracks. 
The PV is required to lie within 60~cm of the geometrical center of the detector along the beam axis. 
The data include a total of 203586 events (186596 in the 
$ee$ category and 16990 in the $\gamma\gamma$ category) that 
satisfy these selection criteria and with the invariant mass of the two EM
clusters $M_{ee/\gamma\gamma}>60$~GeV.

All Monte Carlo samples used in this analysis were generated using {\sc pythia}~\cite{pythia} with CTEQ6L1~\cite{cteq} 
parton distribution functions, and processed through a {\sc geant}-based~\cite{geant} simulation 
of the D0 detector and the same reconstruction software as the data. 
KK graviton signals in the $ee$ and $\gamma\gamma$ decay 
channels are simulated
over the range of parameters $220 \leq M_1 \leq 1050$~GeV and $0.01<k/\bar{M}_{\rm Pl}<0.1$.
The accuracy of the PV association has been studied in KK graviton
events, where the PV reconstruction efficiency is $\approx 98\%$, with $\approx 96\%$ ($\approx 93\%$)
probability to match the true vertex in the $ee$ ($\gamma\gamma$) channel. 
The simulated and observed invariant mass spectra are compared in $ee$ and $\gamma\gamma$
categories separately.
The dominant irreducible 
background in the $ee$ final state is due to the Drell-Yan (DY) process, where 
an $ee$ mass-dependent k factor~\cite{nnlo} is applied to correct the {\sc pythia} spectrum for 
next-to-next-to-leading order effects. 
The dominant irreducible background in the $\gamma\gamma$ final state is
SM $\gamma\gamma$ production, where {\sc pythia} $\gamma\gamma$ events are reweighted to reproduce
the $\gamma\gamma$ invariant mass spectrum predicted by the  
next-to-leading-order calculation of {\sc diphox}~\cite{diphox}. D0 has measured the
SM $\gamma\gamma$ differential cross section with respect to the $\gamma\gamma$ invariant mass,
and in the range used for this analysis (above $60$~GeV) the shape of this distribution from
{\sc diphox} agrees with the data~\cite{gammagamma}.
The leading systematic uncertainty on this background's shape
arises from the choices in the scales used in the {\sc diphox} calculation, and is at the level of
10\%.
The main instrumental background comes from the misidentification of one or two jets as
electrons or photons. The shape of the invariant mass spectrum of this source of events is 
estimated from data by selecting events with EM clusters that are not consistent with electron 
or photon showers using the $\chi^2$ test ($ee$ category) or 
the neural network discriminant ($\gamma\gamma$ category).
Other SM backgrounds, due to DY $\tau\tau$, $W+\gamma$, $WW$, $ZZ$, $WZ$, 
$W+$jets, and $t\bar{t}$ production, are small and 
are estimated using {\sc pythia} Monte Carlo events corrected to account for higher order
 effects~\cite{diboson,wgamma,ttbar}. 

Having obtained the shapes of the invariant mass spectra of the various background sources, 
the background normalization is determined by fitting the invariant mass spectrum of the data
to a superposition of the backgrounds in a low-mass control region 
($60 <M_{ee/\gamma\gamma}< 200$~GeV), where KK gravitons have been excluded 
at the 95\% C.L. by previous searches.
In the fit, the total number of background events is fixed to the number of events observed in the data, and 
the contributions from SM $\gamma\gamma$, DY $ee$, and instrumental background are free 
parameters, while the other SM backgrounds are normalized to their theoretical cross sections.
The fit is performed for the $ee$ and $\gamma\gamma$ categories separately. By varying 
the criteria to select the instrumental background sample and the fitting range, the uncertainty 
of the background normalization procedure is estimated at the level of 2\% (10\%) in 
the $ee$ ($\gamma\gamma$) category.

\begin{figure}
\subfigure{
  \includegraphics[width=0.5\textwidth]{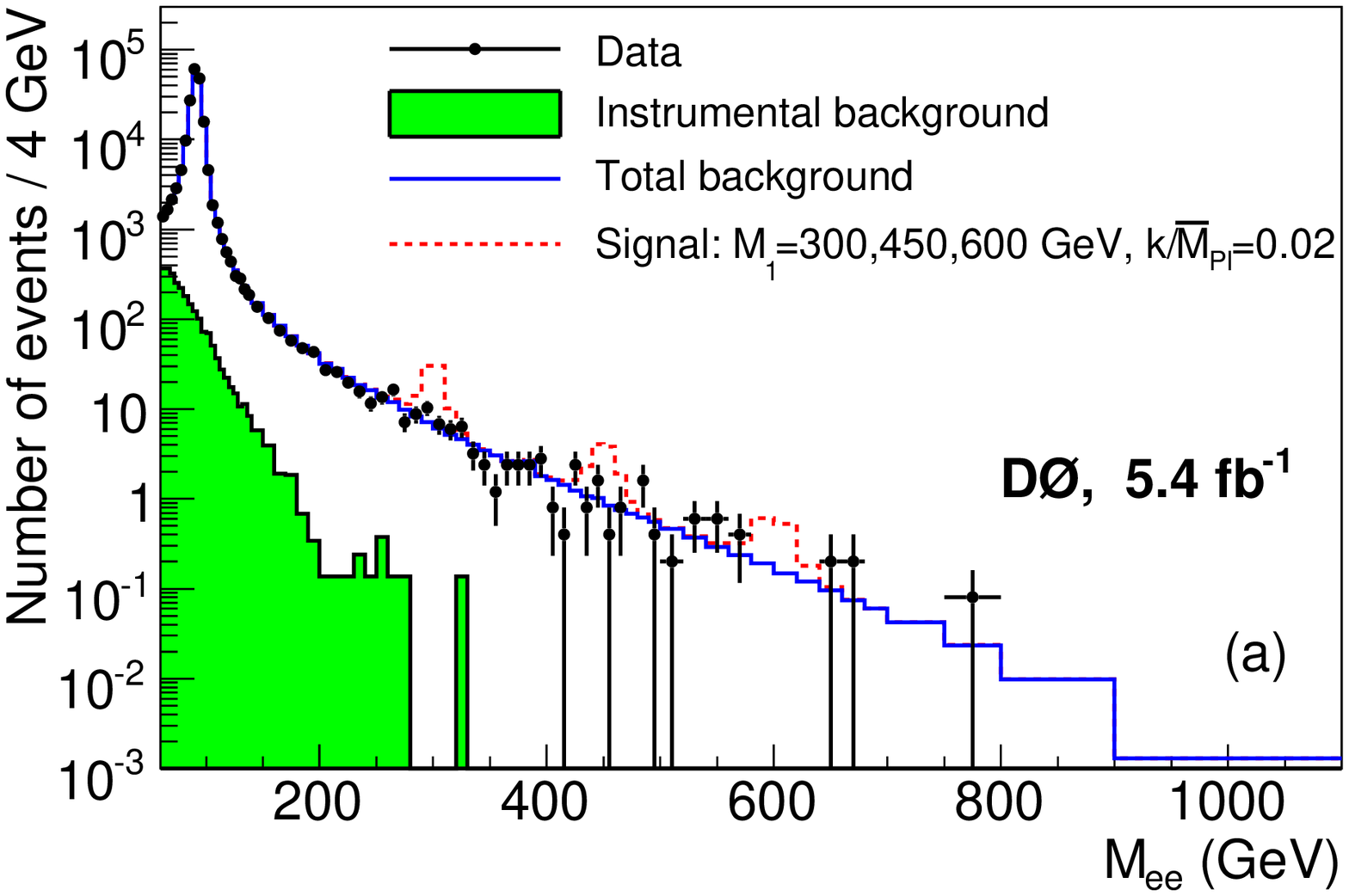}
}
\subfigure{
  \includegraphics[width=0.5\textwidth]{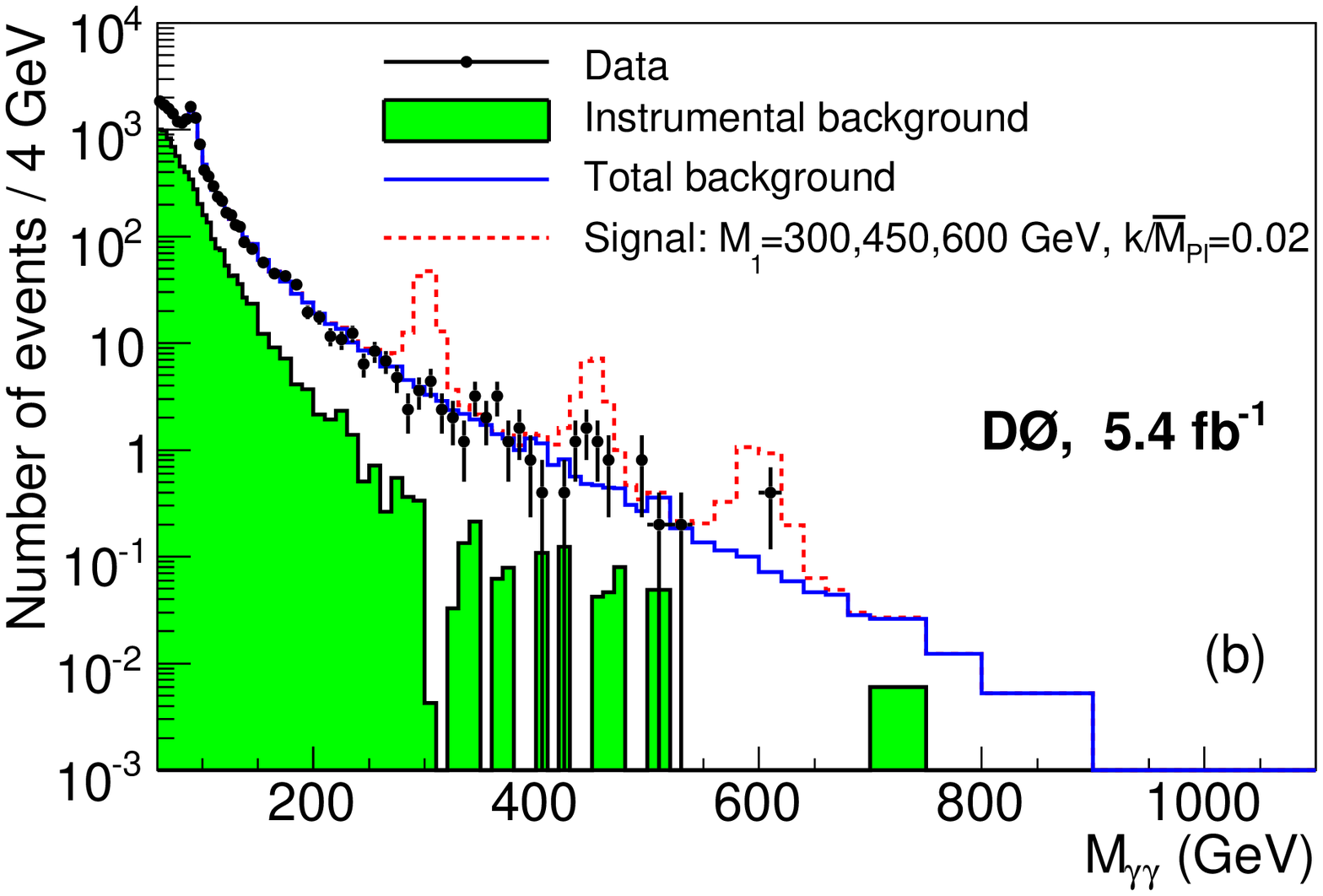}
}
\caption{\label{fig:comparison}Invariant mass spectrum from (a) $ee$ and (b) $\gamma\gamma$ data (points). 
Superimposed are the fitted 
total background shape from SM processes including instrumental background (open histogram) 
and the fitted contribution from events with misidentified clusters alone (shaded histogram). The open 
histogram with dashed line shows the signals
expected from KK gravitons with $M_1=300$, $450$, $600$~GeV (from 
left to right) and 
$k/\bar{M}_{\rm Pl}=0.02$ on top of the total background. 
Invariant masses below $200$~GeV are taken as the control region.}
\end{figure}

Figure~\ref{fig:comparison} shows 
the measured $ee$ and $\gamma\gamma$ invariant mass spectra from the data, superimposed
on the expected backgrounds. 
The data and predicted background are generally in good agreement.
In the region around $450$~GeV there is an excess of events in the $\gamma\gamma$ invariant
mass spectrum. 
As estimated with pseudoexperiments, the probability that this
excess is exclusively due to backgrounds' fluctuations is 0.011, implying
that the background-only hypothesis is disfavored at 
the 2.30 standard deviations (s.d.) level. If we assume that this excess
is due to a KK graviton, including the $ee$ channel reduces the significance to 2.16 s.d..

In the absence of any significant signal for a heavy narrow resonance, we compute upper limits for the 
production cross section of KK gravitons times the branching fraction into the $ee$ final
state using a Poisson log-likelihood ratio (LLR) test~\cite{collie}.
Invariant mass distributions are utilized in the limit calculation.
The $ee$ and $\gamma\gamma$ categories are treated as two independent channels, and then
the two separate LLRs are added to obtain a combined exclusion limit
assuming the 1:2 ratio of the branching fractions. 

Systematic uncertainties on the backgrounds' predictions and on the 
signal efficiency are considered to calculate limits. These include
the integrated luminosity (6.1\%), parton distribution functions
(0.7\% - 6.6\% for the acceptance and 9.2\% - 16.9\% for the graviton production cross section), 
electron and photon identification
efficiency (3.0\% per object), EM cluster energy resolution (6\%), and trigger efficiency (0.1\%). 
The uncertainty on the acceptance due to initial state radiation (ISR) is estimated to be 4\% by
varying the parameters governing ISR in {\sc pythia}.
Uncertainties affecting the expected backgrounds arise from
electron and photon identification efficiency (3.0\% per object), 
mass dependence of the DY $ee$ next-to-next-to-leading order k factor (5.0\%), 
shape of the SM $\gamma\gamma$ invariant mass spectrum, and background normalization.
For the EM energy resolution, the SM $\gamma\gamma$ invariant mass spectrum and the background
normalization we consider both the effects on the normalization and on the shape of the invariant
mass distribution used in extracting limits. For all other systematic sources we consider only
changes to the overall background normalization or signal detection efficiency.
Systematic uncertainties are incorporated via convolution of the Poisson probability
distributions for signal and background with Gaussian distributions corresponding to the different 
sources of systematic uncertainty. Correlations in the systematic uncertainties 
between signal and background in $ee$ and $\gamma\gamma$ categories are taken into account.

\begin{figure}
\includegraphics[width=0.5\textwidth]{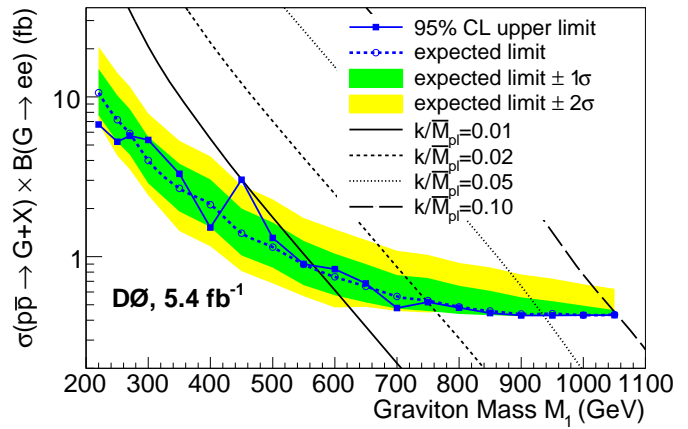}
\caption{\label{fig:limit1} 95\% C.L. upper limit on 
$\sigma(p\bar p \rightarrow G+X)\times B(G\rightarrow ee)$ 
from 5.4 $\rm fb^{-1}$ of integrated luminosity compared with the expected limit and 
the theoretical predictions for different couplings $k/\bar{M}_{\rm Pl}$.}
\end{figure}

\begin{figure}
\includegraphics[width=0.5\textwidth]{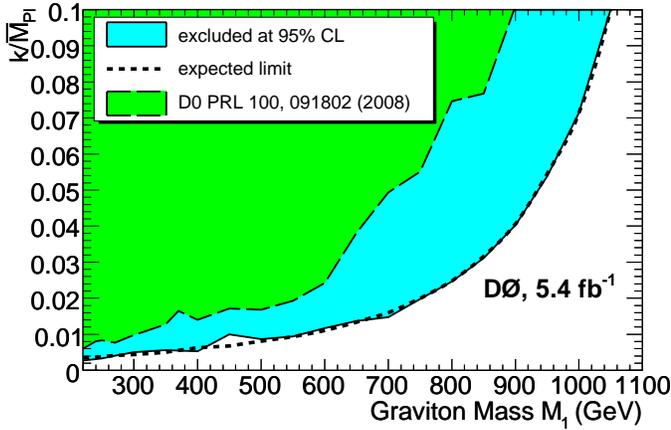}
\caption{\label{fig:limit2} 95\% C.L. upper limit on 
$k/\bar{M}_{\rm Pl}$ versus the graviton mass $M_1$ from 5.4 $\rm fb^{-1}$ 
of integrated luminosity compared with the expected limit 
and the previously published exclusion contour~\cite{d0result}.}
\end{figure}

The resulting limits on the production cross section times
branching fraction into electron-positron pairs of the lightest KK graviton,
$\sigma(p\bar p \rightarrow G+X)\times B(G\rightarrow ee)$,
are given in Table~\ref{tab:limit} and displayed in Fig.~\ref{fig:limit1}. 
As shown in Fig.~\ref{fig:limit2}, using the cross section predictions from the 
Randall-Sundrum model with a k factor of 1.54~\cite{Gkfactor}, we can express these results
as upper limits on the coupling $k/\bar{M}_{\rm Pl}$ as a function of $M_1$. 

\begin{table}[bt]
\caption{95\% C.L. upper limit on $\sigma(p\bar p \rightarrow G+X)\times B(G\rightarrow ee)$ and coupling $k/\bar{M}_{\rm Pl}$ from 5.4 $\rm fb^{-1}$ of integrated luminosity. }
\begin{ruledtabular}
  \begin{tabular}{c|cc|cc}
    Graviton Mass & \multicolumn{2}{c|}{$\sigma \times B(G \rightarrow ee)$ (fb)} & \multicolumn{2}{c}{Coupling $k/\bar{M}_{\rm Pl}$} \\ 
    GeV & Expected & Observed & Expected & Observed \\ \hline
    220 & 10.62 & 6.71 & 0.0034 & 0.0027 \\
    250 & 7.18 & 5.23 & 0.0038 & 0.0033 \\
    270 & 5.91 & 5.69 & 0.0042 & 0.0041 \\
    300 & 4.00 & 5.37 & 0.0044 & 0.0050 \\
    350 & 2.67 & 3.30 & 0.0051 & 0.0056 \\
    400 & 2.12 & 1.52 & 0.0062 & 0.0053 \\
    450 & 1.40 & 3.03 & 0.0068 & 0.0099 \\
    500 & 1.15 & 1.31 & 0.0081 & 0.0087 \\
    550 & 0.89& 0.90& 0.0093 & 0.0094 \\
    600 & 0.75& 0.84& 0.0111  & 0.0117 \\
    650 & 0.65& 0.68& 0.0133  & 0.0136 \\
    700 & 0.56& 0.48& 0.0160  & 0.0147 \\
    750 & 0.53& 0.52& 0.0199  & 0.0197 \\
    800 & 0.48& 0.48& 0.0248  & 0.0247 \\
    850 & 0.46& 0.44& 0.0316  & 0.0312\\
    900 & 0.44& 0.43& 0.0406  & 0.0403\\
    950 & 0.44& 0.43& 0.0545  & 0.0539\\
    1000& 0.43& 0.43& 0.0713  & 0.0713\\
    1050& 0.43& 0.43& 0.0969  & 0.0964\\

  \end{tabular}
\end{ruledtabular}
\label{tab:limit}
\end{table}

In summary, using 5.4 $\rm fb^{-1}$ of integrated luminosity collected with the D0 detector at the
Fermilab Tevatron Collider, we have searched for a heavy narrow resonance in the $ee$ and 
$\gamma\gamma$ invariant mass spectra. The observed spectra agree
with predictions from SM background processes. For the Randall-Sundrum model 
with a warped extra dimension, we set 95\% C.L. upper limits on 
$\sigma(p\bar p \rightarrow G+X)\times B(G\rightarrow ee)$
of the lightest Kaluza-Klein mode of the graviton between 6.7 fb and 0.43 fb for masses between $220$
and $1050$~GeV at the 95\% C.L., which translate into lower limits on the mass $M_1$ of the 
lightest Kaluza-Klein excitation of the graviton between $560$ and $1050$~GeV for couplings of the graviton 
to the SM fields $0.01 \leq k/\bar{M}_{\rm Pl} \leq 0.1$. 
These results represent the most sensitive limits to date.

%
We thank the staffs at Fermilab and collaborating institutions,
and acknowledge support from the
DOE and NSF (USA);
CEA and CNRS/IN2P3 (France);
FASI, Rosatom and RFBR (Russia);
CNPq, FAPERJ, FAPESP and FUNDUNESP (Brazil);
DAE and DST (India);
Colciencias (Colombia);
CONACyT (Mexico);
KRF and KOSEF (Korea);
CONICET and UBACyT (Argentina);
FOM (The Netherlands);
STFC and the Royal Society (United Kingdom);
MSMT and GACR (Czech Republic);
CRC Program and NSERC (Canada);
BMBF and DFG (Germany);
SFI (Ireland);
The Swedish Research Council (Sweden);
and
CAS and CNSF (China).
%

%

\end{document}